\documentclass[aps,pra,superscriptaddress,twocolumn]{revtex4-2}
\usepackage{amsmath}
\usepackage{mathtools}
\usepackage{graphicx}	
\usepackage{bm} 
\usepackage{color}
\usepackage{setspace}
\usepackage{placeins}
\usepackage{natbib}
\usepackage{amssymb}
\begin{document}
\title{Relation between the noise correlations and the spin structure factor for Mott-insulating states in SU$(N)$ Hubbard models}
\date{\today}
\author{Mathias~Mikkelsen}
\email[]{mathias-mikkelsen@phys.kindai.ac.jp}
\author{Ippei~Danshita}
\email[]{danshita@phys.kindai.ac.jp}
\affiliation{Department of Physics, Kindai University, Higashi-Osaka City, Osaka 577-8502, Japan}
\begin{abstract} 
It is well established that the noise correlations measured by time-of-flight imaging in cold-atom experiments, which correspond to the density-density correlations in the momentum space of trapped atomic gases, can probe the spin structure factor deep in the Mott-insulating regime of SU(2) Hubbard models. We explicitly derive the mathematical relation between the noise correlations and the spin structure factor in the strong-interaction limit of SU$(N)$ Hubbard models at any integer filling $\rho$. By calculating the ground states of one-dimensional SU$(N)$ Fermi-Hubbard models for $2\leq N\leq 6$ with use of the density-matrix renormalization-group method, we confirm the relation numerically in the regime of strong interactions 
$U \gg t$, where $U$ and $t$ denote the onsite interaction and the hopping energy. We show that the deviation between the actual noise correlations and those obtained from the spin structure factor scales as approximately $(t/U)^2$ for $\rho=1$ at intermediate and large lattice sizes on the basis of numeric and semi-analytic arguments.
\end{abstract}

\maketitle
\section{Introduction}
\label{sec:section1}

The SU$(N)$ symmetry group plays an important role in many areas of physics \cite{Greiner1994}, with SU(2) and SU(3) being particularly relevant for quantum electrodynamics and chromodynamics respectively \cite{Schwartz2013}. More recently, the experimental realization of fermionic SU$(N)$ symmetry in optical-lattice systems loaded with ultracold atoms, utilizing the nuclear spin degrees of freedom of alkaline-earth(-like) atoms \cite{Taie2012,Hofrichter2016,Ozawa2018,Taie2022,Tusi2022}, has led to renewed relevance of prior investigations into the many-body properties of SU$(N)$ Hubbard and spin models. For the SU$(N)$ Fermi-Hubbard model, the equilibrium phase diagram has seen a number of theoretical investigations \cite{Assaraf1999,Honerkamp2004,Szirmai2005,Buchta2007,Zhao2007,Szirmai2008,Manmana2011,Zhou2014,
Capponi2016,Nie2017,IGP2021}. In the Mott insulating regime of strong repulsive interactions for commensurate integer fillings, the charge gap is open and the low-energy sector of the Hubbard Hamiltonian is described by the SU$(N)$ Heisenberg spin model. In this regime, previous theoretical studies
have predicted a variety of magnetic phases, such as unconventional N\'eel ordered phases~\cite{Toth2010,Yamamoto2020}, dimerized phases~\cite{Buchta2007,Affleck1988,Marston1989}, plaquette ordered phases~\cite{Corboz2012,Corboz2013,Lang2013}, and the coexistence phase of the dimer and N\'eel orders~\cite{Corboz2011}.

In general, observing magnetism inside the Mott-insulating states requires the cooling down to very low temperatures. Very recently, the Mott insulating states for $N=6$ have been successfully cooled down to temperatures as low as $T\simeq 0.1 t/ k_{\rm B}$, where $t$ is the hopping energy, due to the Pomeranchuk cooling mechanism \cite{Taie2022}. Since such a temperature is low enough for magnetic correlations to start developing over a long distance \cite{Bonnes2012}, it is now important to measure long-range spin-spin correlations in experiments. Spin-spin correlation functions at an arbitrary distance have been measured in the case of $N=2$ with alkali-metal atoms \cite{Mazurenko2017} by means of quantum-gas microscope techniques, which allow for addressing the atom number at a single-site resolution. However, while considerable effort has been made to develop similar techniques for alkaline-earth(-like) atoms \cite{Yamamoto2016,Yamamoto2017,Miranda2017,Okuno2020}, it is still difficult to measure spin-spin correlation functions in the Mott-insulating states.

An alternative way to access long-range spin-spin correlations is to analyze the density-density correlations in momentum space. It has been shown that they are closely related to the spin structure factor, which is a Fourier transform of the spin-spin correlation function, in the Mott limit for SU(2) Hubbard models \cite{Altman2004}. The density-density correlations in the momentum space of a trapped ultracold gas can be obtained by measuring the noise correlations of the atom density after a time-of-flight expansion following the release of the gas from the trap~\cite{Folling2005,Greiner2005, Spielman2007,Simon2011,Wurz2018}.
While several papers have made reference to a more general relation between the spin structure factor and the noise correlations in the SU$(N)$ Hubbard model in passing, e.g., Refs.~\cite{Hermele2009,Gorshkov2010,Xu2018}, citing Ref.~\cite{Altman2004}, none of them have explicitly shown this mathematically. In this paper we demonstrate the mathematical relation between the two quantities for Mott states at arbitrary integer filling $\rho$ and investigate the validity of the formula for finite interactions, away from the Mott limit, analytically and numerically. The numerical calculations away from the perfect Mott-limit focus on the SU$(N)$ Fermi-Hubbard model with $2\leq N\leq6$ in one dimension, which we simulate utilizing density-renormalization group-theory (DMRG)\cite{White1992}.

The paper is structured as follows. In Sec.~\ref{sec:section2} we introduce the SU$(N)$ Hubbard model and define the SU$(N)$ spin operators in terms of the creation and annihilation operators. We then introduce the spin structure factor and the noise correlations. In Sec.~\ref{sec:section3} we derive the mathematical relation between the noise correlations and the spin structure factor for the Mott-insulating states in the strong-interaction limit. We then outline the schematic corrections expected at finite interactions from perturbation theory. In Sec.~\ref{sec:section4} we investigate the one-dimensional SU$(N)$ Fermi-Hubbard model for $2 \leq N\leq 6$ numerically using DMRG theory. We investigate strong interactions for which the relation is expected to hold and smaller interactions where it is expected not to hold. We also probe the deviation as a function of the interaction strength and compare it with that obtained by the perturbation theory. In Appendix \ref{sec:AppendixA} we give a more thorough exposition of the second order degenerate perturbation theory. In Appendix \ref{sec:AppendixB} we outline how the numerical results were obtained utilizing the ITensor library.

\section{Model and relevant correlation functions} 
\label{sec:section2}
We consider the SU$(N)$-symmetric Hubbard model on a hypercubic lattice with $L^d$ sites, where $d$ corresponds to the spatial dimension. The Hamiltonian is given by
\begin{align}
\hat{H} = \sum_{\mathclap{\alpha,\langle\bm{l},\bm{m}\rangle}} -t \left(\hat{c}_{\alpha,\bm{l}}^\dagger \hat{c}_{\alpha,\bm{m}} +\text{H.c.}\right)+\frac{U}{2} \sum_{\alpha,\beta ,\bm{l}} \hat{c}_{\alpha,\bm{l}}^\dagger \hat{c}_{\beta,\bm{l}}^\dagger \hat{c}_{\beta,\bm{l}} \hat{c}_{\alpha,\bm{l}}
\label{eq:SUNHubbard}
\end{align} 
where $\langle\bm{l},\bm{m}\rangle$ denotes the sum over nearest neighbors and the indices correspond to the $d$-dimensional index $\bm{l}=(l_1,...,l_d)$. The SU$(N)$ Hamiltonian has $N$ flavors, each with their own set of creation and annihilation operators $\hat{c}_{\alpha,\bm{l}}^\dagger$ and $\hat{c}_{\alpha,\bm{l}}$, where $\alpha$ denotes the flavor. The commutation relations for the creation and annihilation operators determine whether the system is fermionic or bosonic. Note that we consider both fermionic and bosonic cases for analytical calculations shown mainly in Sec.~\ref{sec:section3} whereas numerical simulations in Sec.~\ref{sec:section4} are presented only for the fermionic case, which is directly relevant to experiments with alkaline-earth(-like) atoms in optical lattices. In both cases the Hamiltonian commutes with SU$(N)$ rotations generated by the $N^2-1$ linearly independent generators
\begin{align}
\hat{S}^A_{\bm{l}} = \sum_{\alpha,\beta} \hat{c}_{\alpha,\bm{l}}^\dagger T^A_{\alpha,\beta}\hat{c}_{\beta,\bm{l}},
\label{eq:SUNcartanbasis}
\end{align}
where $T_{\alpha,\beta}^A$ correspond to the matrix elements of the fundamental matrix representation of the SU$(N)$ generators defined by the commutation relation
\begin{align}
[\hat{T}^A,\hat{T}^B]=i f^{AB}_C \hat{T}^C.
\end{align}  
Here $f^{AB}_C $ is a structure constant, which in the case of SU$(2)$ is the fully antisymmetric Levi-Civita symbol $\epsilon^{AB}_C$. We employ the normalization  $\text{Tr}(\hat{T}^A \hat{T}^B)=\delta_{AB}/2$. These generators can be understood as generalized spin operators in analogy with the usual definition of spin in SU(2) systems.

A useful way to probe the spin ordering of the Hubbard model is the spin correlation function which measures the spatial spin correlations as  
\begin{align}
 S(\bm{l},\bm{m}) = \sum_{A} \langle\hat{S}^A_{\bm{l}} \hat{S}^A_{\bm{m}} \rangle. 
\end{align}
Using the relation 
\begin{align}
\sum_{A}T^A_{\alpha,\beta}T^A_{\gamma,\delta} = \frac{1}{2}(\delta_{\alpha\delta}\delta_{\beta\gamma}-\frac{1}{N}\delta_{\alpha \beta} \delta_{\gamma \delta})
\end{align}
the spin correlation function can be written in terms of the particle operators as 
\begin{align}
\sum_{A}\hat{S}^A_{\bm{l}} \hat{S}^A_{\bm{m}} \!=\! \frac{1}{2}\left[\sum_{\alpha,\beta} \hat{c}_{\alpha,\bm{l}}^\dagger \hat{c}_{\beta,\bm{l}}  \hat{c}_{\beta,\bm{m}}^\dagger \hat{c}_{\alpha,\bm{m}} \!-\! \sum_{\alpha,\beta}\frac{1}{N}\hat{n}_{\alpha,\bm{l}}\hat{n}_{\beta,\bm{m}}\right].
\label{eq:spinspincartan}
\end{align}
For SU$(N)$-symmetric states, $\langle\hat{S}^A_{\bm{l}} \hat{S}^A_{\bm{m}} \rangle = \langle \hat{S}^B_{\bm{l}} \hat{S}^B_{\bm{m}} \rangle $ is satisfied for all $A$ and $B$. The spin-spin correlations can then be fully determined from the diagonal spin operators and the full spin-spin correlation function can be found from the simple expression (using any two different flavors $\alpha$ and $\beta$) 
\begin{align}
 \sum_{A} \langle\hat{S}^A_{\bm{l}} \hat{S}^A_{\bm{m}} \rangle = \frac{N^2-1}{2}[\langle \hat{n}_{\alpha,\bm{l}} \hat{n}_{\alpha,\bm{m}}\rangle - \langle \hat{n}_{\alpha,\bm{l}} \hat{n}_{\beta\neq \alpha,\bm{m}}\rangle].
\label{eq:spinspinSUN1}
\end{align}

The spin structure factor is defined as the Fourier transform of the spin-spin correlation function
\begin{align}
\tilde{S}(\bm{k})=\frac{1}{L^d}\sum_{\bm{l},\bm{m}}  e^{i(\bm{l}-\bm{m})\cdot \bm{k}} \sum_{A} \langle \hat{S}^A_{\bm{l}} \hat{S}^A_{\bm{m}} \rangle.
\end{align} 
Here $\bm{k}=(k_1,..,k_d)=\left(\frac{2 n_1 \pi}{L},...,\frac{2 n_d \pi}{L}\right)$, where $n_1,...,n_d$ are integers, are the reciprocal lattice momentum vectors.

The spin-spin correlations are straightforward to calculate numerically using Eq.~(\ref{eq:spinspinSUN1}). While this quantity has been measured  experimentally in the case of $N=2$ by means of the quantum-gas microscope techniques for alkali-metal atoms~\cite{Mazurenko2017}, it is not the case for $N>2$, which requires the use of alkaline-earth(-like) atoms, thus far.
Accessing the density-density correlations in the momentum space of a trapped ultracold gas, however, is more straightforward by measuring the noise correlations of the atom density after a standard time-of-flight expansion of the gas \cite{Altman2004,Folling2005,Greiner2005,Spielman2007,Simon2011,Wurz2018}. Based on the explicit relation between the two quantities, for convenience, we hereafter call the density-density correlations in the momentum space the noise correlations. The full noise correlations are given as a sum over the flavor-resolved noise correlations, 
\begin{align}
G(\bm{k},\bm{k}')&= \sum_{\alpha,\beta}G_{\alpha,\beta}(\bm{k},\bm{k}'),
\end{align}   
with the latter defined as 
\begin{align}
G_{\alpha,\beta}(\bm{k},\bm{k}')= \langle \hat{n}_{\alpha,\bm{k}} \hat{n}_{\beta,\bm{k}'} \rangle - \langle \hat{n}_{\alpha,\bm{k}} \rangle \langle \hat{n}_{\beta,\bm{k}'} \rangle,
\end{align}  
where $\bm{k}$ and $\bm{k}'$ are the reciprocal lattice momentum vectors. These can also be calculated in terms of the Fourier transform of the real-space one-body density matrix and four-point correlation function
\begin{align}
&G_{\alpha,\beta}(\bm{k},\bm{k}')= \nonumber  \\
& \frac{1}{L^{2d}}\sum_{\mathclap{\bm{l},\bm{l}',\bm{m},\bm{m}'}}e^{i(\bm{l}-\bm{l}')\cdot \bm{k}+i(\bm{m}-\bm{m}') \cdot \bm{k}'} \langle \hat{c}_{\alpha,\bm{l}}^\dagger \hat{c}_{\alpha,\bm{l}'} \hat{c}_{\beta,\bm{m}}^\dagger  \hat{c}_{\beta,\bm{m}'}  \rangle -  \nonumber \\
&\frac{1}{L^{2d}}\sum_{\mathclap{\bm{l},\bm{l}',\bm{m},\bm{m}'}}e^{i(\bm{l}-\bm{l}')\cdot \bm{k}+i(\bm{m}-\bm{m}')\cdot \bm{k}'} \langle \hat{c}_{\alpha,\bm{l}}^\dagger \hat{c}_{\alpha,\bm{l}'} \rangle \langle \hat{c}_{\beta,\bm{m}}^\dagger \hat{c}_{\beta,\bm{m}'} \rangle.
\label{eq:fullmomentumnoisecorrelations}
\end{align}

While the noise correlations are experimentally accessible, the evaluation of four-point correlation functions is more challenging numerically than the simple two-point correlations required for the spin-spin correlation function. 

\section{Analytic results}
\label{sec:section3}
\subsection{Noise correlations and the spin structure factor in the Mott limit}
\label{sec:section3.1}
For the Mott-insulating states corresponding to $\rho/N$ filling (where $\rho =1,2,\ldots,N-1$), a relation between the noise correlations and the spin structure factor can be derived. We consider a lattice consisting of $L^d$ sites, with $d$ corresponding to the spatial dimension.  Any Mott state can be written as a linear combination of the restricted set of Fock states $|\psi \rangle=\sum_\mu c_{\mu} |\mu \rangle$ that fulfill the constraints $\sum_{\bm{l}} \langle \mu | \hat{n}_{\alpha,\bm{l}}| \mu \rangle =\rho \frac{L^d}{N}$ and $\sum_{\alpha} \langle \mu | \hat{n}_{\alpha,\bm{l}} | \mu \rangle=\rho$. Note that this also implies that  $\sum_{\bm{l}} \langle \psi | \hat{n}_{\alpha,\bm{l}}| \psi \rangle = \rho \frac{L^d}{N}$ and $\sum_{\alpha} \langle \psi | \hat{n}_{\alpha,\bm{l}} | \psi \rangle= \rho$. Angular brackets will be used as a shorthand for expectation values with respect to a Mott state $| \psi \rangle$ in the rest of this section. These constraints are sufficient to derive the relation and are obeyed by both fermionic and bosonic Mott states. As the effective spin Hamiltonian for the two cases is different, the spin structure factor will display different physics, but it will always be related to the noise correlations by the formula we derive below. For expectation values with respect to a Mott state, the one-body density matrix can only have nonzero contributions when $\bm{l}=\bm{m}$ (other terms would connect to Fock states outside the restricted set), i.e., 
\begin{align}
\langle \hat{c}^\dagger_{\alpha,\bm{l}} \hat{c}_{\alpha,\bm{m}} \rangle = \langle \hat{n}_{\alpha,\bm{l}}  \rangle \delta_{\bm{l},\bm{m}} ,
\end{align} 
while the four-point correlation function can only have contributions when  $\bm{l}=\bm{l}'$ and $\bm{m}=\bm{m}'$ or  $\bm{l}=\bm{m}'$ and $\bm{m}=\bm{l}'$, i.e. 
\begin{align}
\langle \hat{c}^\dagger_{\alpha,\bm{l}} \hat{c}_{\alpha,\bm{l}'} \hat{c}^\dagger_{\beta,\bm{m}}  \hat{c}_{\beta,\bm{m}'}  \rangle =
& \langle   \hat{n}_{\alpha,\bm{l}} \hat{n}_{\beta,\bm{m}} \rangle  \delta_{\bm{l},\bm{l}'}\delta_{\bm{m},\bm{m}'}[1-\delta_{\bm{l},\bm{m}}] \nonumber  \\
&+\delta_{\bm{l},\bm{m}'} \delta_{\bm{m},\bm{l}'}\langle \hat{c}^\dagger_{\alpha,\bm{l}} \hat{c}_{\alpha,\bm{m}} \hat{c}^\dagger_{\beta,\bm{m}}  \hat{c}_{\beta,\bm{l}}  \rangle.
\label{eq:4points}
\end{align}
Here the factor $1-\delta_{\bm{l},\bm{m}}$ on the first term prevents double counting of the $\bm{l}=\bm{l}'=\bm{m}=\bm{m}'$ case. Some of these terms (depending on the integers $\rho$ and $N$) will be zero for fermions due to the Pauli-exclusion principle, but this is not important for the following derivation. In order to relate the noise correlations to the spin structure factor, rearranging the ordering of the particle operators is necessary.  Utilizing the bosonic or fermionic commutation relations we can rewrite Eq.~(\ref{eq:4points}) as
\begin{align}
\langle \hat{c}^\dagger_{\alpha,\bm{l}} \hat{c}_{\alpha,\bm{l}'} \hat{c}^\dagger_{\beta,\bm{m}}  \hat{c}_{\beta,\bm{m}'}  \rangle &= \langle   \hat{n}_{\alpha,\bm{l}} \hat{n}_{\beta,\bm{m}} \rangle  \delta_{\bm{l},\bm{l}'}\delta_{\bm{m},\bm{m}'}[1-\delta_{\bm{l},\bm{m}}] \nonumber \\
 &+ \delta_{\bm{l},\bm{m}'} \delta_{\bm{m},\bm{l}'}[\eta \langle\hat{c}^\dagger_{\alpha,\bm{l}} \hat{c}_{\beta,\bm{l}} \hat{c}^\dagger_{\beta,\bm{m}}  \hat{c}_{\alpha,\bm{m}} \rangle \nonumber \\
 &-\eta \langle \hat{n}_{\alpha,\bm{l}} \rangle \delta_{\bm{l},\bm{m}}+\langle\hat{n}_{\alpha,\bm{l}}\rangle \delta_{\alpha,\beta}],
\label{eq:4points2}
\end{align}
where $\eta =1$ for bosons and $\eta=-1$ for fermions. To evaluate $G(\bm{k},\bm{k}')$, we sum over the flavors $\sum_{\alpha,\beta}$. From Eq.~(\ref{eq:spinspincartan}) this means that the second term on the right-hand side of Eq.~(\ref{eq:4points2}) is related to the spin-spin correlations as
\begin{align}
&\sum_{\alpha,\beta}\langle\hat{c}^\dagger_{\alpha,\bm{l}} \hat{c}_{\beta,\bm{l}} \hat{c}^\dagger_{\beta,\bm{m}}  \hat{c}_{\alpha,\bm{m}} \rangle  
\nonumber\\
&\,\,\,\, =   2 \sum_{A} \langle \hat{S}^A_{\bm{l}} \hat{S}^A_{\bm{m}} \rangle +\frac{1}{N}\sum_{\alpha,\beta} \langle   \hat{n}_{\alpha,\bm{l}} \hat{n}_{\beta,\bm{m}} \rangle.
\end{align}

Evaluating all the Fourier transforms in Eq.~(\ref{eq:fullmomentumnoisecorrelations}), we find
\begin{align}
G_{\text{spin}}(\bm{k},\bm{k}')&= \eta \rho \left(\frac{\rho}{N}+\eta\right)\delta_{\bm{k},\bm{k}'} \nonumber \\
&+\frac{\rho}{L^d}\left[-\eta N-\rho+\frac{2 \eta}{\rho} \tilde{S}(\bm{k}-\bm{k}')\right].
\label{eq:momentumspinrelation}
\end{align}
As we want to quantitatively probe this relation away from the ideal Mott limit, we label the noise correlations obtained via this formula $G_{\text{spin}}(\bm{k},\bm{k}')$ in order to distinguish them from the noise correlations obtained by directly evaluating Eq.~(\ref{eq:fullmomentumnoisecorrelations}), which can have contributions beyond the Mott sector for finite $U/t$. The SU$(N)$-independent contribution to the $\delta$ function $\eta^2 \rho=\rho$ at $\bm{k}-\bm{k}'=0$ can be removed by starting with a normal-ordered four-point correlation function which corresponds to subtracting $\sum_{\alpha} \delta_{\bm{k},\bm{k}'}\langle n_{\alpha,\bm{k}}\rangle$ from the noise correlations. Doing this more clearly elucidates the effect of $N$ at $\bm{k}-\bm{k}'=0$ when we compare the numerical calculations. We will therefore compare $\tilde{G}(\bm{k},\bm{k}')=G(\bm{k},\bm{k}')-\sum_{\alpha} \delta_{\bm{k},\bm{k}'}\langle n_{\alpha,\bm{k}}\rangle$ with 
\begin{align}
\tilde{G}_{\text{spin}}(\bm{k},\bm{k}')=& \eta\frac{\rho^2}{N}\delta_{\bm{k},\bm{k}'}+\frac{\rho}{L^d}\left[-\eta N-\rho+\frac{2 \eta}{\rho} \tilde{S}(\bm{k}-\bm{k}')\right].
\label{eq:momentumspinrelation2}
\end{align}

If we invert Eq.~(\ref{eq:momentumspinrelation}) and insert the relation between the density distribution after a time-of-flight experiment and the momentum distribution in the initial state of a trapped atomic gas \cite{Altman2004}, the spin structure factor can be determined in terms of the experimentally measurable density noise correlation $G_{\text{TOF}}(\bm{r}-\bm{r}')$, where $\bm{r}$ and $\bm{r}'$ correspond to the spatial coordinates after the time of flight. The formula is given by 
\begin{align}
\tilde{S}\biggl(\bm{Q}(\bm{r})-\bm{Q}(\bm{r}')\biggr)&=\frac{\rho L^d}{2}\biggl[\frac{\eta}{\rho W^{2d}}G_{\text{TOF}}(\bm{r}-\bm{r}')\biggr. \nonumber \\ 
&\biggl. -\left(\frac{\rho}{N}+\eta\right)\delta_{\bm{Q}(\bm{r}),\bm{Q}(\bm{r}')}+\frac{1}{L^d}(N+\eta \rho)\biggr],
\label{eq:experimentalspin}
\end{align} 
where $W=\hbar t/(a_0 m)$, with $a_0$ corresponding to the width of the Wannier state on the lattice and $m$ to the mass of the particles. Here $\bm{Q}(\bm{r})=m\bm{r}/ (\hbar t)$ describes the correspondence between the real-space location after the time of flight and the lattice momentum in the initial state. 

\subsection{Deviation from the perfect Mott limit}
While the above formula is exact in the Mott limit, experiments are performed at a finite value of $U/t$ and it is therefore important to understand how the true noise correlations deviate from the result obtained from the spin structure factor as the interaction strength is lowered. For this purpose, we investigate the absolute difference $\delta G_\text{spin}(\bm{k},\bm{k}')=|G(\bm{k},\bm{k}')-G_{\text{spin}}(\bm{k},\bm{k}')|$. From the application of second-order degenerate perturbation theory (Appendix \ref{sec:AppendixA}), we can show that this difference will have contributions proportional to  $(t/U)^n$, where $n$ is a positive integer. Ignoring terms with $n>2$, it can be written as 
\begin{align}
&\delta G_{\text{spin}}(\bm{k},\bm{k}')= \bigg|\frac{t}{U} \delta G_{\text{spin},1}(\bm{k},\bm{k}')+\left(\frac{t}{U}\right)^2 \delta G_{\text{spin},2}(\bm{k},\bm{k}')\bigg|
\label{eq:nonMottcorrection}
\end{align}
where
\begin{align}
&\delta G_{\text{spin},1}(\bm{k},\bm{k}')=\nonumber \\
&  2\text{Re} \biggl(\,\,\,\,\,\,\,\,\,\sum_{\mathclap{\alpha,\beta,\gamma,\langle \bm{l},\bm{m}\rangle}}\,\,\,\,\, \,\,\, 
\langle\psi^{U}_{\text{Mott}}|\hat{n}_{\alpha,\bm{k}} \hat{n}_{\beta,\bm{k}'} \hat{\mathbb{K}}_{\gamma,\bm{l},\bm{m}}|\psi^{U}_0 \Biggr. \rangle\nonumber \\
&\,\,\,\,\,\,\,\,\,\,\,\,\,\,\,\,\,-\rho \sum_{\mathclap{\alpha,\gamma,\langle \bm{l},\bm{m}\rangle}}\,\,\,\,\,\,  \Bigl[\langle \psi^{U}_{\text{Mott}}|\hat{n}_{\alpha,\bm{k}}\hat{\mathbb{K}}_{\gamma,\bm{l},\bm{m}}|\psi^{U}_0 \rangle \Bigr. \nonumber \\
&\,\,\,\,\,\,\,\,\,\,\,\,\,\,\,\,\,+\biggl. \Bigl.\langle \psi^{U}_{\text{Mott}}|\hat{n}_{\alpha,\bm{k}'}\hat{\mathbb{K}}_{\gamma,\bm{l},\bm{m}}|\psi^{U}_0 \rangle \Bigr] \biggr) 
\label{eq:1stordercontribution}
\end{align}
and
\begin{align}
&\delta G_{\text{spin},2}(\bm{k},\bm{k}')= \nonumber \\
&\sum_{\mathclap{\alpha,\beta,\gamma,\langle \bm{l},\bm{m}\rangle,\langle\bm{l}',\bm{m}'\rangle}} \,\,\,\,\,\,\,\,\,\,\,\,\,\,\,\,\,\langle\psi^{U}_0| \hat{\mathbb{K}}_{\gamma,\bm{l},\bm{m}}^\dagger \hat{n}_{\alpha,\bm{k}} \hat{n}_{\beta,\bm{k}'} \hat{\mathbb{K}}_{\gamma,\bm{l}',\bm{m}'}|\psi^{U}_0\rangle  \nonumber  \\
&-4\text{Re}\biggl(\,\,\,\,\,\,\,\, \sum_{\mathclap{\alpha,\gamma,\langle \bm{l},\bm{m}\rangle}}\,\,\,\,\,\,\,\langle \psi^{U}_{\text{Mott}}|\hat{n}_{\alpha,\bm{k}}\hat{\mathbb{K}}_{\gamma,\bm{l},\bm{m}}|\psi^{U}_0 \rangle \biggr)\nonumber \times \\
&\text{Re}\biggl(\,\,\,\,\,\,\,\sum_{\mathclap{\alpha,\gamma,\langle \bm{l},\bm{m}\rangle}}\,\,\,\,\,\,\,\langle \psi^{U}_{\text{Mott}}|\hat{n}_{\alpha,\bm{k}'}\hat{\mathbb{K}}_{\gamma,\bm{l},\bm{m}}|\psi^{U}_0 \rangle\biggr).
\label{eq:2ndordercontribution}
\end{align}
Here $|\psi^{U}_0\rangle$ corresponds to the ground state of the effective Hamiltonian (Heisenberg model) at the value $U$ in units of $t$, while $|\psi^{U}_{\text{Mott}} \rangle$ contains the correction to the ground state from excited states of the effective Hamiltonian and 
\begin{align}
\hat{\mathbb{K}}_{\gamma,\bm{l},\bm{m}}= \left(\hat{c}_{\gamma,\bm{l}}^\dagger \hat{c}_{\gamma,\bm{m}} +\text{H.c.}\right)
\end{align}
corresponds to the nearest-neighbor hopping terms for flavor $\gamma$.

As we show in Appendix \ref{sec:AppendixA}, the first-order term can be simplified as many of the matrix elements cancel out, but it is difficult to obtain a universal, simple, and analytic expression for SU$(N)$. In the next section, we will discuss the corrections in more detail based on numerical simulations. 

\section{Numerical simulations}
\label{sec:section4}
In this section, we investigate the one-dimensional SU$(N)$ Fermi-Hubbard model, evaluating the ground-state properties with use of the DMRG algorithm \cite{White1992} implemented with the ITensor library \cite{Fishman2020}. One way to simulate the SU$(N)$ Fermi-Hubbard model utilizes a lattice size equivalent to the physical size of the system with a large local Hilbert space containing all possible single-site occupations similar to Ref.\cite{Manmana2011}. Another strategy, which we employ in our simulations, considers a larger effective lattice size but a smaller local Hilbert space, instead representing different flavors by sublattices necessitating beyond-nearest-neighbor hopping, similar to Ref.\cite{Tusi2022}. See Appendix \ref{sec:AppendixB} for a more detailed breakdown of our simulation method.

All calculations in this section are for a fixed equal number of particles of each flavor. The effective lattice size utilized in the calculation scales with $N$ and the inclusion of beyond-nearest-neighbor hopping terms means that the required bond dimension scales quickly with the lattice size. The ground-state properties are well converged for a bond dimension of 400 for SU(2)-SU(5) and 800 for the SU(6) simulations. Our method is not that practical for elaborate SU(5) and SU(6) simulations as the achievable system sizes, that can be calculated in a reasonable time frame are relatively small ($L=15$ and $18$ in our simulations), but the results are nevertheless useful for checking the validity of the relation between the spin structure factor and the noise correlations. For fermions $\eta=-1$, which according to Eq.(\ref{eq:momentumspinrelation}) means that peaks in the spin structure factor will be observed as antipeaks in the noise correlations.

\subsection{Strong interactions - Mott-like states}
\label{sec:section4.1}
The integer Mott states can be observed for commensurate lattice sizes $L=N,2N, ...$ and strong repulsion $U/t$. In order to investigate the relation derived in Sec. \ref{sec:section3.1}, we calculate $\tilde{G}(k-k')$  directly utilizing Eq.~(\ref{eq:fullmomentumnoisecorrelations}) and compare it with $\tilde{G}_{\text{spin}}(k-k')$  of Eq.~(\ref{eq:momentumspinrelation2}), where the spin structure factor is calculated based on Eq.~(\ref{eq:spinspinSUN1}). To illustrate that the formula works for various combinations of $N$ and $\rho$, we investigate the fermionic case of Eq.~(\ref{eq:SUNHubbard}) for $2\leq N\leq 6$ and at different fillings. As $\rho=m$ and $\rho=N-m$ display the same physics due to particle-hole symmetry \cite{Dufour2015}, however, we only show plots for $\rho\leq N/2$. 

\begin{figure*}
\centering
\includegraphics[width=1\linewidth]{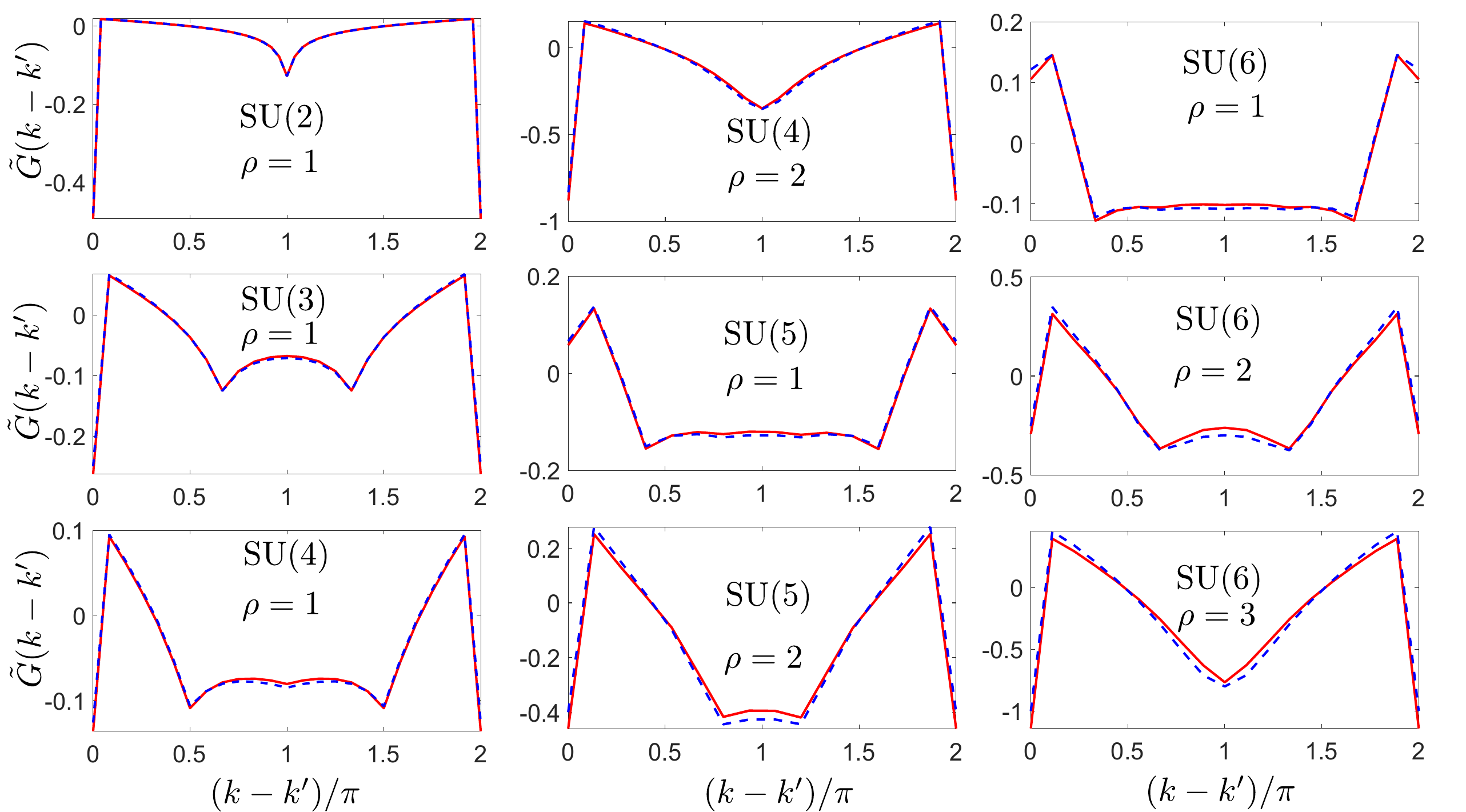}
\caption{Plot of $\tilde{G}(k-k')$ calculated directly through the noise correlations (\ref{eq:fullmomentumnoisecorrelations}) (red solid line) and $\tilde{G}_{\text{spin}}(k-k')$ obtained from the spin structure factor using Eq.~(\ref{eq:momentumspinrelation2}) (blue striped line) for the ground state of the $U/t=20$ Fermi-Hubbard M.model The SU(2) calculations are done at $L=50$, SU(3) and SU(4) at $L=24$, SU(5) at $L=15$, and SU(6) at $L=18$.}
\label{fig:U20results} 
\end{figure*}

These plots are shown in Fig.~\ref{fig:U20results}. Details about the system size for each model are contained in the captions. Visually, it is clear that the two calculations give similar results. The small difference can be attributable to the fact that the system is not in the perfect Mott limit, even at $U/t=20$. The difference is largest for $\rho>1$ in the SU(5) and SU(6) models, suggesting that a larger $U/t$ is required for the ground state to become Mott-like in these cases. 

The main result of this section is the approximate confirmation of Eq.~(\ref{eq:momentumspinrelation2}) at large $U/t$. The physics for different $N$ and $\rho$ has been thoroughly investigated in terms of the Heisenberg model \cite{Dufour2015}, which describes the charge-gapped Mott limit well. Here we briefly summarize the features of the noise correlations and hence the spin structure factor (see Ref.~\cite{Dufour2015} for more details). For $\rho=1$ or $N-1$, the Heisenberg model predicts $N-1$ gapless SU$(N)$ modes resulting in sharp peaks located at $k=\frac{2}{N} \pi$ and $\frac{2(N-1)}{N} \pi$ in the spin structure factor, which we observe as anti-peaks. For $N=5$ and $\rho=2$, a gapless phase is also expected. These antipeaks are harder to discern, but are still present in our calculations. For $\frac{N}{\rho}=2$ in the SU(4) and SU(6) cases, a gapped dimerized phase is expected and the spin structure factor displays a broadened peak at $k=\pi$ consistent with faster off-diagonal decay at longer distance (exponential decay is expected at long distances in larger systems). For $N=6$ and $\rho=2$, the system is also gapped, but due to the small system size in our calculations, the anti-peak looks as sharp as in the gapless cases. In fact, it was shown in Ref.~\cite{Dufour2015} that a system size on the order of $L=50$ was required to change the concavity of the peak. The location of the antipeaks in all cases correspond to $2 k_F$ where $k_F$ is the Fermi momentum \cite{Capponi2016}.

\subsection{Weaker interactions}
\label{sec:section4.2}
While the formula seems relatively accurate for $U/t=20$, a more thorough investigation as a function of the interaction strength is required. Towards this end we first investigate the qualitative features of the noise correlation and spin structure factor for a smaller value of $U/t=5$, where the formula is not expected to hold in general, following up with a quantitative investigation of the discrepancy between  $G(k-k')$ and $G_{\text{spin}}(k-k')$  as a function of $U/t$.

\begin{figure*}
\centering
\includegraphics[width=1\linewidth]{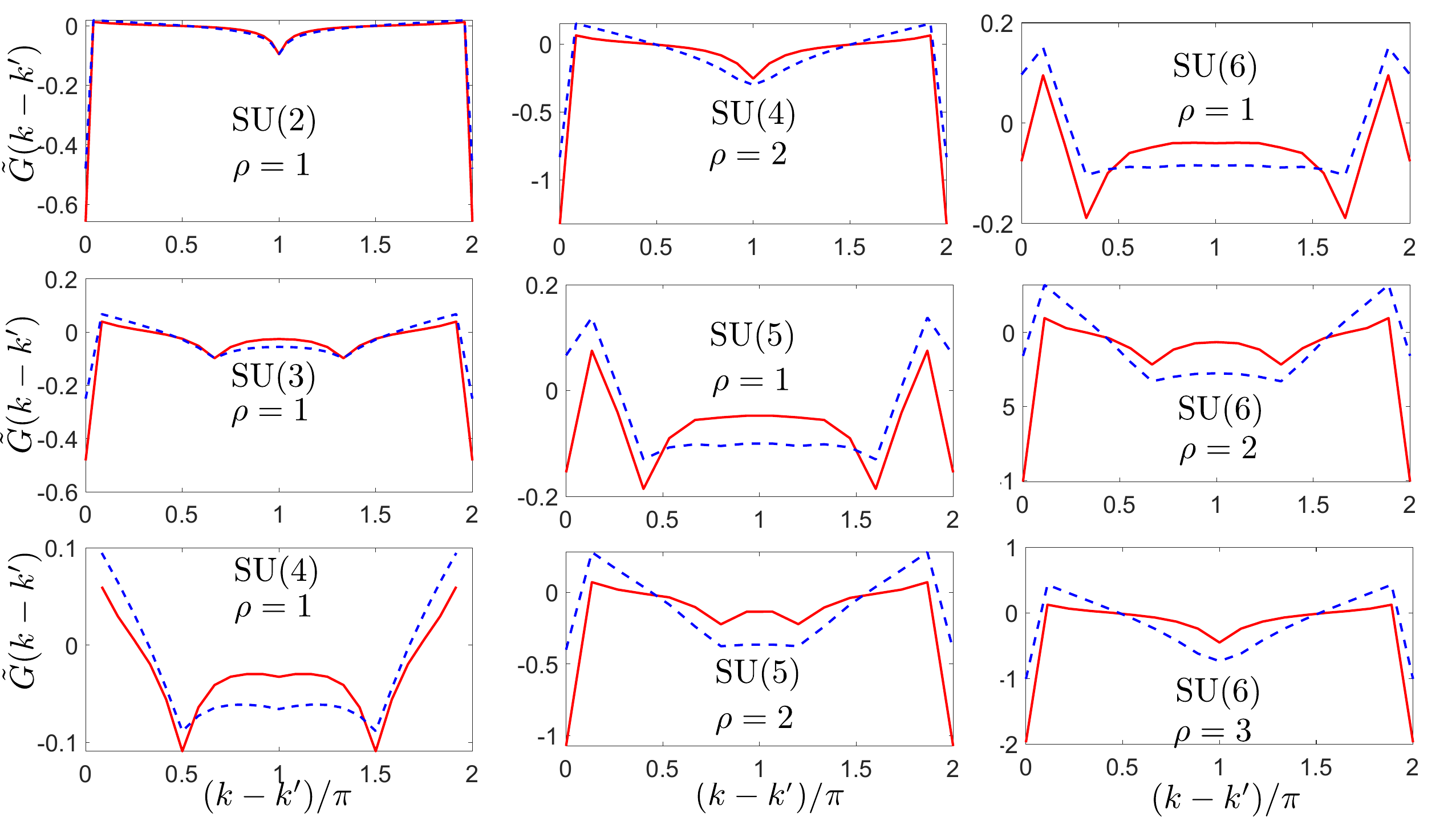}
\caption{Plot of $\tilde{G}(k-k')$ calculated directly through the noise correlations (\ref{eq:fullmomentumnoisecorrelations}) (red solid line) and $\tilde{G}_{\text{spin}}(k-k')$ obtained from the spin structure factor using Eq.~(\ref{eq:momentumspinrelation2}) (blue striped line) for the ground state of the $U/t=5$ Fermi-Hubbard model. The SU(2) calculations are done at $L=50$, SU(3) and SU(4) at $L=24$, SU(5) at $L=15$, and SU(6) at $L=18$.}
\label{fig:U5results} 
\end{figure*}

For $N=2$, a Mott-insulating state is the ground state for any positive $U$ in the thermodynamic limit \cite{Capponi2016}. For $N>2$ and $\rho=1$ the system is a metallic $N$-component Luttinger liquid with one gapless charge mode and $N-1$ gapless SU$(N)$ modes below a finite critical interaction $U_{\rm C}$ \cite{Assaraf1999,Manmana2011,Capponi2016}.  For $U>U_{\rm C}$ the system displays the same phases as in the Mott limit. The transition is predicted to be of the Kosterlitz-Thoules type and we therefore do not expect to discern a sharp distinction between the two for the lattice sizes considered in this paper. At half-filling a charge density wave phase of $N$-mers is predicted for $U<0$ with a Kosterlitz-Thouless type of transition at $U=0$ to a gapped dimerized phase for $U>0$ with an exponentially slow opening of the charge gap \cite{Buchta2007}.  In Fig.~\ref{fig:U5results} we plot the same as in Fig.~\ref{fig:U20results}, but for a relatively small interaction $U/t=5$, where we expect the Mott-limit relation to no longer be applicable. It is clear that the effective interaction strength is decreased as $N$ is increased, that is, for $N=2$ and 3 the system is still relatively close to the Mott-limit results, while the results are substantially different for $N>3$. This is consistent with previous investigations \cite{Assaraf1999,Manmana2011,Capponi2016} which suggest that the critical value of $U_{\rm C}$ increases with $N$. In general, Eq.~(\ref{eq:momentumspinrelation2}) is no longer quantitatively accurate; however, both $\tilde{G}(k-k')$  and the  $\tilde{G}_{\text{spin}}(k-k')$ display antipeaks at the same location (twice the Fermi momentum). The antipeaks are much more pronounced for  $\tilde{G}(k-k')$, which in all cases display sharper peaks. This includes the $N/ \rho = 2$ case where  $\tilde{G}_{\text{spin}}(k-k')$ still displays a much broader anti-peak. 

Here $\tilde{G}_{\text{spin}}(k-k')$, which is based on the spin structure factor, only probes the SU$(N)$ excitations, while the full noise correlations also probe the charge excitation and the general narrowing of the antipeaks may be related to the closing of the charge gap at smaller interactions. Indeed, even the half-filling case which is charge gapped at any $U>0$ in the thermodynamic limit is essentially (charge) gapless at small $U$ at the considered system size due to the exponentially slow opening of the gap. 

In order to better understand how valid Eq.~(\ref{eq:momentumspinrelation2}) is for weaker interactions, we need to quantify how the full noise correlations, which include the charge sector, differ from the pure spin sector contribution calculated via the spin structure factor. To do this, we investigate the absolute difference $\delta G_\text{spin}(k,k')=|G(k,k')-G_{\text{spin}}(k,k')|$, averaging the values away from $\delta_{k,k'}$, that is, 
\begin{equation}
\Delta G_{\text{spin}}=\frac{1}{L-1}\,\,\,\,\,\,\,\,\,\,\,\,\sum_{\mathclap{(k-k')=2\pi/L}}^{\mathclap{2\pi(L-1)/L}}
\,\,\,\,\,\,\,\,\,\,\,\,\delta G_{\text{spin}}(k-k').
\end{equation}

\begin{figure*}
\centering
\includegraphics[width=1\linewidth]{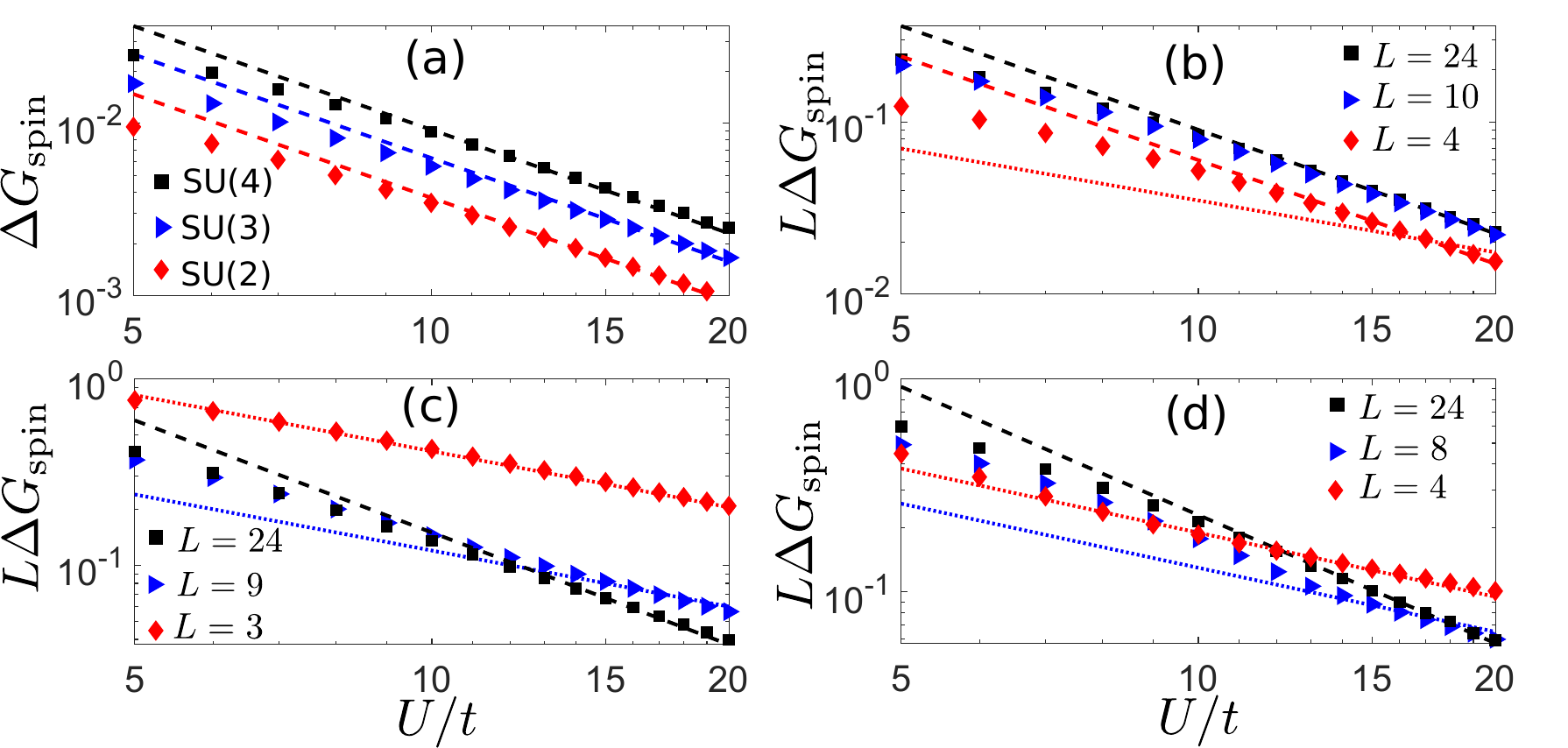}
\caption{(a) Plot of $\Delta G_{\text{spin}}$ as a function of $U$ for $N=2, 3,$ and $4$ at $L=24$. (b)-(d) Plots of $L \Delta G_{\text{spin}}$ as a function of $U$ for different lattice sizes: (b) SU(2) for $L=4,10,24$, (c) SU(3) for $L=3,9,24$, and (d) SU(4) for $L=4,8,24$. The dashed lines are proportional to $(t/U)^2$, while the dotted lines are proportional to $t/U$.}
\label{fig:Averagedeviation} 
\end{figure*}

To simplify the required calculations and discussion, we restrict ourselves to $\rho=1$. In Fig.~\ref{fig:Averagedeviation}(a) we plot $\Delta G_{\text{spin}}$ as a function of $U/t$ for $N=2$, 3, and 4, where $L=24$ (we choose the same system size to make the most accurate comparison). The deviation is well approximated by $(t/U)^2$ for all $N$ in $U/t\in [10,20]$. The deviation for smaller values of $U/t$ grows somewhat slower, but it is clear that at $U/t=5$ (corresponding to Fig.~\ref{fig:U5results}), the deviation is significant, with more than an order of magnitude difference from the results at $U/t=20$. However, the proportionality constant grows larger with $N$, consistent with interactions being effectively weaker. The small proportionality constant for SU(2) is the reason why Eq.~(\ref{eq:momentumspinrelation2}) still gives relatively good agreement with the full noise correlations for SU(2).

The most striking aspect of this result, however, is the apparent vanishing of the first-order contribution, with the second-order contribution dominating the deviation. In Appendix \ref{sec:AppendixA} we analyze the first-order correction and analytically show that many of the matrix elements cancel out. For identical bosons, one can analytically show that the contribution of the first-order terms scales as $1/L$, while numerical evaluation of the relevant matrix elements suggests that it entirely disappears for SU(2) fermions. In Figs.~\ref{fig:Averagedeviation}(b)-(d) we plot the deviation for $N=2,3$, and $4$ at different system sizes, scaling the results with $L$ to make the sizes comparable. We see that the $(t/U)^2$ scaling indeed holds at small system sizes for SU(2), consistent with the vanishing of the first-order terms. For $N>2$, the numerical results suggest that the situation is similar to identical bosons, with smaller system size leading to a scaling closer to $t/U$, but the second-order contribution dominating at larger system sizes, because the first-order contribution decays faster with the system size $L$. 

For intermediate lattice sizes and finite values within experimental range, which are roughly $L\gtrsim 20$ and $U/t\gtrsim 10$~\cite{Taie2022}, the second-order contribution is therefore the most important. The $(t/U)^2$ scaling of the deviation is highly specific to the noise correlations and results in Eq.~(\ref{eq:momentumspinrelation2}) being more accurate at finite $U/t$ than one would naively expect.

\section{conclusion}
\label{sec:section5}
We have investigated the relation between the noise correlations and the spin structure factor for SU$(N)$ models, deriving an exact relation for Mott states at arbitrary integer filling on a square lattice in any dimension for bosons and fermions.
We have investigated this relation numerically in one-dimensional SU$(N)$ Fermi-Hubbard models for $2\leq N\leq6$. These results suggest that the formula is reasonably accurate in all cases for large interaction $U/t$ and we have determined that the expected deviation for finite interaction $U/t$ at intermediate and large system sizes scales as $(t/U)^2$, with the first-order error being negligible. Our results are relevant for experimentally probing SU$(N)$ magnetism in currently available cold atom experiments as they suggest that the spin structure factor can be determined through time-of-flight measurements for realistic finite interactions $U/t$.

\begin{acknowledgments}
The calculations were performed utilizing the ITensor c++ library \cite{Fishman2020}. We would also like to acknowledge D. Kagamihara, R. Kaneko, S. Taie, and Y. Takahashi for useful discussions and comments. This work was financially supported by JSPS KAKENHI (Grants No. JP18H05228, JP21H01014, and JP22K14007), by MEXT Q-LEAP (Grant No. JP-MXS0118069021), and by JST FOREST (Grant No. JPMJFR202T).

\end{acknowledgments}

\begin{appendix}
\section{Pertubative analysis of the error as a function of $t/U$}
\label{sec:AppendixA}
We write the Hamiltonian as 
\begin{align}
\hat{H} &= \hat{H}_0+\hat{\mathbb{K}}, \\
\hat{H}_0 &= \frac{U}{2} \sum_{\alpha,\beta ,\bm{l}} \hat{c}_{\alpha,\bm{l}}^\dagger \hat{c}_{\beta,\bm{l}}^\dagger \hat{c}_{\beta,\bm{l}} \hat{c}_{\alpha,\bm{l}}, \\
 \hat{\mathbb{K}} &= -t\sum_{\mathclap{\gamma,\langle\bm{l},\bm{m}\rangle}}  \left(\hat{c}_{\gamma,\bm{l}}^\dagger \hat{c}_{\gamma,\bm{m}} +{\rm \text{H.c.}}\right)=-t\sum_{\mathclap{\gamma,\langle\bm{l},\bm{m}\rangle}} \hat{\mathbb{K}}_{\gamma,\bm{l},\bm{m}},
\end{align} 
considering the hopping terms $\hat{T}$ as a perturbation on top of $\hat{H}_0$. The ground-state manifold exactly corresponds to the Mott states as defined in Sec. \ref{sec:section3.1}. It is well known that the effective Hamiltonian obtained from second-order degenerate perturbation theory is the SU$(N)$ Heisenberg model, but in this appendix we are interested in the state correction which can be used to calculate the correction to a given observable (in our case the noise correlations). The ground-state correction for a system where degeneracy is lifted at second order is given by 
\begin{align}
|\psi_0 \rangle &= |\psi^{U}_0 \rangle+\sum_{J\neq 0} a_{0,J}^{(1)}|\psi^{U}_J \rangle\nonumber +\\
&\sum_{\psi_j \in D} \frac{\langle \phi_j | \hat{T}| \psi^{U}_0 \rangle}{\langle \phi_j|\hat{H}^{(0)}|\phi_j \rangle -\langle  \psi^{U}_0  |\hat{H}^{(0)}| \psi^{U}_0 \rangle }|\phi_{j} \rangle .
\label{eq:PertubationState}
\end{align}
Here $|\psi^{U}_0 \rangle$ is the ground state of the effective Hamiltonian, while $|\psi^{U}_J \rangle$ are exited states of the effective Hamiltonian and $| \phi_j \rangle$ describes the states in the complement to the ground-state manifold $D$, i.e., the non Mott states. The coefficients in the second term are not important for our calculation, but can be found in standard course materials for degenerate perturbation theory, for example Ref.~\cite{Zwiebach2018}. Indeed, as the first two terms describe the part of the ground state which is within the Mott-restricted region and is captured by Eq.~(\ref{eq:momentumspinrelation}), the expectation value with respect to $|\psi^{U}_{\text{Mott}} \rangle = |\psi^{U}_0 \rangle+\sum_{J\neq0} a_{I,J}^{(1)}|\psi^{U}_J \rangle$ should correspond to the result obtained by Eq.~(\ref{eq:momentumspinrelation}). What we focus on here is the deviation from this.

It is clear that the hopping terms connect $|\psi^{U}_0 \rangle$ only to states that differ from the Mott state (which has a total energy $L \rho(\rho-1)U/2$) by having $\rho-1$ and $\rho+1$ total occupations at a pair of neighboring sites (total energy of $[(L-1) \rho(\rho-1)+(\rho+2)(\rho+1)+(\rho-1)(\rho-2)]U/2$). This means that the energy difference, which appears in the denominator of the third term on the right-hand side of Eq.~(\ref{eq:PertubationState}), is always $-U$. In addition, the intermediate states correspond exactly to those obtained by applying the hopping terms to the ground state of the effective Hamiltonian and we can write the state as 
\begin{align}
|\psi_0 \rangle = |\psi^{U}_{\text{Mott}} \rangle+\frac{t}{U}\sum_{\gamma,\langle\bm{l},\bm{m}\rangle} \hat{\mathbb{K}}_{\gamma,\bm{l},\bm{m}}|\psi^{U}_0 \rangle.
\end{align}

Taking the expectation values of the relevant operators (and noting that the momentum distribution is flat in the Mott limit, i.e., $\sum_{\alpha} \langle \hat{n}_{\alpha,\bm{k}} \rangle = \rho$), we can write the first and second order contributions in $t/U$ to the noise correlations as in Eqs.(\ref{eq:nonMottcorrection})-(\ref{eq:2ndordercontribution}) in the main text.

It is possible to simplify the calculation of these considerably, further using our knowledge of the Mott states. In particular, we will focus on the first-order contribution, which is analytically tractable and the numerical calculations in the main text suggest that it becomes insignificant for many physical situations of interest. To simplify the discussion, we will assume a one-dimensional system, but similar results should be obtainable in three dimensions with the only difference being more hopping terms corresponding to more neighbors. In one dimension, the hopping terms can be written as 
\begin{align}
\sum_{\mathclap{\gamma,\langle l,n\rangle}} \hat{\mathbb{K}}_{\gamma,l,n} &=\sum_{\mathclap{\gamma,\langle l,n\rangle}}[\hat{c}_{\alpha,n}^\dagger \hat{c}_{\alpha,n+1}+\hat{c}_{\alpha,n+1}^\dagger \hat{c}_{\alpha,n}]\delta_{l,n} \nonumber \\
&=\sum_{n} [\hat{c}_{\alpha,n}^\dagger \hat{c}_{\alpha,n+1}+\hat{c}_{\alpha,n+1}^\dagger \hat{c}_{\alpha,n}].
\end{align}
In order to evaluate the first-order correction to the noise correlations, we must evaluate the first-order correction to the one-body density matrix and four-point correlation function. To simplify the notation, we shorten $\langle \psi^{U}_{\text{Mott}}| ... |\psi^{U}_0 \rangle$ to $\langle ... \rangle$ in the following derivation. The one-body density matrix is relatively simple, with the only possible nonzero contributions (the rest would connect to non-Mott states) being for $l=l'+1$ and $l=l'-1$, i.e.
\begin{align}
&\sum_{\alpha}\langle \hat{c}^\dagger_{\alpha,l} \hat{c}_{\alpha,l'} \sum_{n,\gamma} \hat{\mathbb{K}}_{\gamma,n} \rangle = \nonumber \\
&\,\,\,\,\,\,\,\,\,\,\,\,\,\,\,\,\,\,\,\,\sum_{\alpha,\gamma} \left[ \langle \hat{c}^\dagger_{\alpha,l} \hat{c}_{\alpha,l+1} \hat{c}^\dagger_{\gamma,l+1} \hat{c}_{\gamma,l} \rangle \delta_{l,l'-1}\right.
\nonumber \\
&\,\,\,\,\,\,\,\,\,\,\,\,\,\,\,\,\,\,\,\,\,\,\,\,\,\,\,\,\,\,\left.+\langle \hat{c}^\dagger_{\alpha,l+1} \hat{c}_{\alpha,l} \hat{c}^\dagger_{\gamma,l} \hat{c}_{\gamma,l+1} \rangle \delta_{l,l'+1}\right].
\end{align}

For the four-point correlation to give nonzero contributions, we must always pair an annihilation operator at a site with a corresponding creation operator at the same site, although it can be of a different flavor, in order to stay within the restricted Mott space.  There will be two types of possible non-zero contributions. One of these corresponds to having a number operator of a flavor on one site, that is 
\begin{align}
 &\sum_{\alpha=\beta}\langle \hat{c}^\dagger_{\alpha,l} \hat{c}_{\alpha,l'} \hat{c}^\dagger_{\beta,m} \hat{c}_{\beta,m'}  \sum_{n,\gamma} \hat{\mathbb{K}}_{\gamma,n}^\dagger\rangle =  \nonumber \\
 &\sum_{\alpha,\beta,\gamma} \bigg[\langle \hat{c}^\dagger_{\alpha,l} \hat{c}_{\alpha,l+1} \hat{c}^\dagger_{\beta,m} \hat{c}_{\beta,m} \hat{c}^\dagger_{l+1,\gamma} \hat{c}_{l,\gamma} \rangle \delta_{m,m'}\delta_{l,l'-1}+ \nonumber \\
 & \langle \hat{c}^\dagger_{\alpha,l+1} \hat{c}_{\alpha,l} \hat{c}^\dagger_{\beta,m} \hat{c}_{\beta,m} \hat{c}^\dagger_{\gamma,l} \hat{c}_{\gamma,l+1} \rangle \delta_{m,m'}\delta_{l,l'+1} + \nonumber \\
 & \langle \hat{c}^\dagger_{\alpha,l} \hat{c}_{\alpha,l} \hat{c}^\dagger_{\beta,m} \hat{c}_{\beta,m+1} \hat{c}^\dagger_{\gamma,m+1} \hat{c}_{\gamma,m} \rangle \delta_{l,l'}\delta_{m,m'-1}+ \nonumber \\
 & \langle \hat{c}^\dagger_{\alpha,l} \hat{c}_{\alpha,l} \hat{c}^\dagger_{\beta,m+1} \hat{c}_{\beta,m} \hat{c}^\dagger_{\gamma,m} \hat{c}_{\gamma,m+1} \rangle \delta_{l,l'}\delta_{m,m'+1}\bigg]+\dots
\end{align}

These can be simplified, as any Mott state is an eigenstate of the operator $\sum_{\alpha} \hat{n}_{\alpha,l} |\psi_{\text{Mott}}\rangle =\rho |\psi_{\text{Mott}}\rangle$ (with an eigenvalue which is site independent). Using the commutation relations, we can move the number operator in the first two terms to the right-hand side, which results in the extra terms 
\begin{align*}
&\sum_{\alpha,\beta} \bigg( [-\langle \hat{c}^\dagger_{l,\alpha} \hat{c}_{\alpha,l+1} \hat{c}^\dagger_{\beta,l+1} \hat{c}_{\beta,l} \rangle \delta_{m,l} \nonumber \\
&+\langle \hat{c}^\dagger_{\alpha,l} \hat{c}_{\alpha,l+1} \hat{c}^\dagger_{\beta,l+1} \hat{c}_{\beta,l} \rangle \delta_{l+1,m}] \delta_{m,m'}\delta_{l,l'-1}
 + \nonumber \\
&[\langle \hat{c}^\dagger_{\alpha,l+1} \hat{c}_{\alpha,l} \hat{c}^\dagger_{\beta,l} \hat{c}_{\beta,l+1} \rangle \delta_{m,l} \nonumber \\
&-\langle \hat{c}^\dagger_{\alpha,l+1} \hat{c}_{\alpha,l} \hat{c}^\dagger_{\beta,l} \hat{c}_{\beta,l+1} \rangle \delta_{l+1,m}] \delta_{m,m'} \delta_{l,l'+1}\bigg).
\end{align*}
Taking the Fourier transform of these terms results in zero as the positive and negative terms cancel out. Evaluating the number operator, the Fourier transform over the remaining terms can be written as 
\begin{align}
&\frac{\rho}{L}\! \sum_{l,\alpha,\gamma}\!\left(\! e^{-ik} \langle \hat{c}^\dagger_{\alpha,l} \hat{c}_{\alpha,l+1} \hat{c}^\dagger_{\gamma,l+1} \hat{c}_{\gamma,l}\rangle \!+\! e^{ik} \langle \hat{c}^\dagger_{\alpha,l+1} \hat{c}_{\alpha,l} \hat{c}^\dagger_{\gamma,l} \hat{c}_{\gamma,l+1} \rangle\right. \nonumber  \\
&
\left.+ e^{-ik'} \langle \hat{c}^\dagger_{\alpha,l} \hat{c}_{\alpha,l+1} \hat{c}^\dagger_{\gamma,l+1} \hat{c}_{\gamma,l}\rangle + e^{ik'} \langle \hat{c}^\dagger_{\alpha,l+1} \hat{c}_{\alpha,l} \hat{c}^\dagger_{\gamma,l} \hat{c}_{\gamma,l+1} \rangle \right).
\end{align}
This corresponds to the value obtained from the Fourier transform of the two-point correlation function multiplied by $\rho$ and these terms therefore cancel out in Eq.~(\ref{eq:1stordercontribution}) (the calculations so far holds for both bosons and fermions). This means that the only contribution to the first-order correction comes from the second type of contribution to the four-point correlation function, the one which pairs operators of different flavor on the same site. This contribution is given by 
\begin{widetext}
\begin{align}
\frac{1}{L^2} \sum_{l,m} [1-\delta_{l,m}-\delta_{l,m+1}] \sum_{\alpha,\beta,\gamma}&\left( e^{i (m-l)(k'-k)} \left(e^{ik}\langle \hat{c}^\dagger_{\alpha,l+1} \hat{c}_{\alpha,m} \hat{c}^\dagger_{\beta,m} \hat{c}_{\beta,l} \hat{c}^\dagger_{\gamma,l} \hat{c}_{\gamma,l+1} \rangle+e^{-ik'}\langle \hat{c}^\dagger_{\alpha,l} \hat{c}_{\alpha,m} \hat{c}^\dagger_{\beta,m} \hat{c}_{\beta,l+1} \hat{c}^\dagger_{\gamma,l+1} \hat{c}_{\gamma,l} \rangle \right) \right. \nonumber \\ 
&\!\!\!\!\!\!\!\!\!\! \left. +e^{-i(m-l)(k'-k)}\left(e^{ik'}\langle \hat{c}^\dagger_{\alpha,m} \hat{c}_{\alpha,l} \hat{c}^\dagger_{\beta,l+1} \hat{c}_{\beta,m} \hat{c}^\dagger_{\gamma,l} \hat{c}_{\gamma,l+1} \rangle+e^{-ik}\langle \hat{c}^\dagger_{\alpha,m} \hat{c}_{\alpha,l+1} \hat{c}^\dagger_{\beta,l} \hat{c}_{\beta,m} \hat{c}^\dagger_{\gamma,l+1} \hat{c}_{\gamma,l} \rangle \right)\right),
\label{eq:1stordernonzero}
\end{align}
\end{widetext}
where the factor $1-\delta_{l,m}-\delta_{l,m+1}$ takes care of double counting terms that were already counted in the evaluation of the first contribution.

Unlike the other terms, these are dependent on the specific model and filling. All results so far can essentially be applied to the simple one-component Bose-Hubbard model as well (a slight change in Eq.~(\ref{eq:PertubationState}) is required as the Mott-limit ground state is nondegenerate but the end result will have the same type of matrix elements), for which the matrix elements in Eq.~(\ref{eq:1stordernonzero}) can be evaluated analytically in terms of the Mott state $|\rho \rho... \rho\rangle$. This results in a first-order contribution 
\begin{align}
&\delta G_{\text{spin},1}^{\text{Boson}}(k,k')= \nonumber \\ 
&4\rho (2 \rho^2+3 \rho+1)\bigg|\frac{1}{L}[\cos(k)+\cos(k')]-\cos(k) \delta_{k,k'}\bigg|.
\end{align}
Note that the contribution away from $k=k'$ scales as $1/L$ and the first-order contribution is therefore going to become less important as the system size increases.  For $\rho=1$ in the SU(2) case and $\rho=2$ in the SU(4) case, numerical evaluation of the matrix elements using exact diagonalization shows that  

\begin{align*}
&\langle \hat{c}^\dagger_{\alpha,m} \hat{c}_{\alpha,l+1} \hat{c}^\dagger_{\beta,l} \hat{c}_{\beta,m} \hat{c}^\dagger_{\gamma,l+1} \hat{c}_{\gamma,l} \rangle \\
&\,\,\,\,\,\,\,\,\,\,\,\,\,\,= -\langle \hat{c}^\dagger_{\alpha,l+1} \hat{c}_{\alpha,m} \hat{c}^\dagger_{\beta,m} \hat{c}_{\beta,l} \hat{c}^\dagger_{\gamma,l} \hat{c}_{\gamma,l+1} \rangle  
\end{align*}
and
\begin{align*}
&\langle \hat{c}^\dagger_{\alpha,m} \hat{c}_{\alpha,l} \hat{c}^\dagger_{\beta,l+1} \hat{c}_{\beta,m} \hat{c}^\dagger_{\gamma,l} \hat{c}_{\gamma,l+1} \rangle \\
&\,\,\,\,\,\,\,\,\,\,\,\,\,\,=-\langle \hat{c}^\dagger_{\alpha,l} \hat{c}_{\alpha,m} \hat{c}^\dagger_{\beta,m} \hat{c}_{\beta,l+1} \hat{c}^\dagger_{\gamma,l+1} \hat{c}_{\gamma,l} \rangle,
\end{align*}
which results in a purely imaginary contribution to the Fourier transform and therefore a zero-contribution to the first-order term, i.e.,
\begin{align}
&\delta G_{\text{spin},1}^{\text{SU(2)}}(k,k')= 0.
\end{align}
The numerical calculations in the main text supports this, as the deviation is still determined by the second-order term, even for small lattice sizes. For $\rho=1$ in the SU(3) and SU(4) cases, the first-order correction does not vanish, but the calculations presented in the main text indicate that the behavior is similar to that of single-component bosons. That is, the first-order contribution is dominant at smaller lattice sizes, but the second-order contribution becomes dominant at larger lattice sizes, indicating that the first-order contribution scales inversely with the lattice size.

\section{Numerical representation of SU$(N)$ utilizing matrix product states and tensor networks}
\label{sec:AppendixB}
In order to numerically investigate the  SU$(N)$ Fermi-Hubbard model, we utilize the ITensor library. Within the standard library  the representation of spinless fermions and spinful electrons on a lattice is already efficiently implemented. We therefore build on top of this and represent the $N$-component Hubbard model in terms of these building blocks. For even $N$, we utilize the two-component electron representation built into the library with distinct sublattices corresponding to different flavors. For example, the SU(6) model is represented by the one-dimensional Hamiltonian
\begin{align}
\hat{H} &= -t \sum_{j}\left(\hat{c}_{\uparrow,j}^\dagger \hat{c}_{\uparrow,j+3} + \hat{c}_{\downarrow,j}^\dagger \hat{c}_{\downarrow,j+3}  + {\rm \text{H.c.}}\right) \nonumber \\
&+U \sum_{\mu,\nu}\sum_{j=1,4,...}^{3 L-2}\left( \hat{n}_{\mu,j} \hat{n}_{\nu,j+1}+\hat{n}_{\mu,j+1} \hat{n}_{\nu,j+2}+\hat{n}_{\mu,j} \hat{n}_{\nu,j+2}\right) \nonumber \\
&+U\sum_{j,\mu \neq \nu}  \hat{n}_{\mu,j} \hat{n}_{\nu,j}
\end{align} 
where $\mu,\nu=\uparrow,\downarrow$. If we name the sublattices $A, B$, and $C$ the flavors correspond to $A_{\uparrow}, A_{\downarrow}, B_{\uparrow}, B_{\downarrow}, C_{\uparrow}$, and $C_{\downarrow}$. The SU(4) model can be represented in a similar way, but is simpler as only two sublattices are required. 
  
For the SU(3) and SU(5) calculations, we utilize a system of spinless fermions with three sublattices or five sublattices, i.e.  for SU(3) the Hamiltonian is given by 
\begin{align}
\hat{H} &= -t \sum_{j} \hat{c}_{j}^\dagger \hat{c}_{,j+3}  + {\rm \text{H.c.}} \nonumber \\
&+U \sum_{j=1,4,...}^{3 L-2} \left( \hat{n}_{j} \hat{n}_{j+1}+\hat{n}_{j+1} \hat{n}_{j+2}+\hat{n}_{j} \hat{n}_{j+2}\right)
\end{align} 
with the flavors corresponding to the $A, B$, and $C$ sublattices.

In order to calculate the four-point correlation functions  $\langle \hat{c}^\dagger_{\alpha,l} \hat{c}_{\alpha,l'} \hat{c}^\dagger_{\beta,m}  \hat{c}_{\beta,m'}  \rangle$ in the matrix product state representation, an ordering in terms of lattice indices, as the operators are sequentially applied to the state, must be assumed. We therefore split the four-point correlation function into expectation values over different operator sequences where fermionic anti-commutation rules have been applied when rearranging the original sequence in terms of increasing index. 

We consider the normal-ordered four-point correlation function as this minimizes the number of matrix elements that we are required to calculate. Note that the noise correlations can be obtained from the Fourier transform of this function by adding $\delta_{\alpha,\beta} \delta_{k,k'}\langle n_{\alpha,k}\rangle$, as we saw in the main text. We will consider  $P_{lmjk}=\langle \hat{c}_{l}^\dagger \hat{c}_{m}^\dagger \hat{c}_{j}  \hat{c}_{k} \rangle$ for spinless fermions as an example, the correlations can be calculated in exactly the same way for either the $\uparrow$ or $\downarrow$ components of the SU(2) fermions, while the results for mixing $\uparrow$ and $\downarrow$ components is similar, but more nonzero terms are required. The correlation functions of the full SU$(N)$ model are obtained by considering the calculated correlation function on and between the sublattices defined in the above Hamiltonians. For the SU$(N)$ symmetric case, we therefore only need to calculate it for one computational component  $\uparrow$. 

For arbitrary values of $l, m, j,$ and $k$, we can enumerate all the matrix elements required for the calculation of the four-point correlation function in the ITensor library. If all site indices are different $l\neq m\neq j\neq k$ there are 24 possible ways to arrange them, for example.  Note that the number of elements that we are required to calculate reduces drastically due to the properties of the four-point correlation function, namely, 
\begin{align}
P_{lmjk}&=P_{jklm}=P_{mlkj}=P_{kjml}, \\
P_{mljk}=&P_{lmkj}=P_{jkml}=P_{kjlm}=-P_{lmjk},
\end{align}
which means that these eight elements can be obtained by calculating one of them (this means we only have to consider three out of the 24 possible arrangements when $l\neq m\neq j\neq k$, for example). Additionally, $P_{kkmn}=P_{mnkk}=0$, as these involve the sequential application of two creation or annihilation operators to the same site. When all indices or three indices are equal, there is therefore no contribution to the four-point correlation function. In the case of two indices being equal, any term of the form $P_{kkmn}=P_{mnkk}$ is likewise zero. The remaining terms are given by 
\begin{eqnarray}
l=m<j=k:& \quad \quad P_{lkkl}  &= \langle \hat{c}_l^\dagger \hat{c}_l \hat{c}_k^\dagger \hat{c}_k \rangle, \\
l=m<j<k:& \quad \quad P_{jllk}  &= \langle \hat{c}_l^\dagger \hat{c}_l \hat{c}_j^\dagger \hat{c}_k \rangle, \\
l<m<j=k:& \quad \quad P_{ljjm}  &=  \langle \hat{c}_l^\dagger \hat{c}_m \hat{c}_j^\dagger \hat{c}_j \rangle, \\
l<m=j<k:& \quad \quad P_{ljjk}  &= \langle \hat{c}_l^\dagger \hat{c}_j^\dagger \hat{c}_j \hat{c}_k \rangle.
\end{eqnarray}
The first term corresponds to four matrix elements, while the remaining terms correspond to eight matrix elements using the symmetries of the four-point correlation function. 

When all indices are different and assuming $l<m<j<k$, we get the terms
\begin{align}
P_{lmjk}  &=  \langle \hat{c}_l^\dagger \hat{c}_m^\dagger \hat{c}_j  \hat{c}_k \rangle, \\
P_{lkjm}  &=  - \langle \hat{c}_l^\dagger \hat{c}_m \hat{c}_j \hat{c}_k^\dagger \rangle, \\
P_{ljmk}  &=  -\langle \hat{c}_l^\dagger \hat{c}_m \hat{c}_j \hat{c}_k^\dagger \rangle,
\end{align}
all of which correspond to eight matrix elements. We are therefore left with the need to manually write code for evaluating seven expectation values in ITensor (which corresponds to 52 matrix elements due to symmetries). The full correlation function is obtained by iterating over the lattice indices as $\sum_{l=1}^N\sum_{m=l}^N\sum_{j=m}^N\sum_{k=j}^N$. Note that the number of expectation values required grows proportionally to $L^4$ and the numeric costs therefore grows relatively quickly with the lattice size.

In order to implement this in ITensor, the correct implementation of Jordan-Wigner strings is also necessary as the fermionic operators are implemented as hardcore bosonic operators.  Overall, the implementation is somewhat involved and to ensure the validity of our results we check that the four-point correlations are identical to those obtained from exact diagonalization calculations in small systems  (for exact diagonalization calculations it is simpler to represent the SU$(N)$ system in terms of sequential lattices corresponding to each component).

\end{appendix}

\bibliography{library}

\end{document}